\documentstyle[preprint,prl,aps,floats,epsfig]{revtex}
\begin{document}
\tighten
\draft
\preprint{nucl-th/9502032}
\title{Glauber isovector spin responses for ($\vec{p},\vec{n}$) reactions at
494 MeV}
\author{A. De Pace}
\address{Istituto Nazionale di Fisica Nucleare, Sezione di Torino, via Giuria
1,
I-10125 Torino, Italy}
\date{February 13, 1995}
\maketitle
\begin{abstract}
The separated isovector spin-longitudinal and spin-transverse responses from a
recent ($\vec{p},\vec{n}$) quasifree experiment at 494 MeV are analyzed in a
Glauber theory framework up to two-step contributions. Nuclear correlations are
treated in the continuum random phase approximation, accounting for the
spreading width of the particle-hole states. A good description of the
spin-longitudinal response is achieved, hinting at some evidence of pionic
effects in the quasielastic region. The spin-transverse response is
underestimated and we show that multistep contributions, while sizable, are not
sufficient to provide an explanation of the discrepancy.
\end{abstract}

In recent experiments \cite{McC92,Che93,Tad94} complete sets of polarization
transfer coefficients have been measured for quasifree ($\vec{p},\vec{n}$)
scattering from $^2$H, $^{12}$C and $^{40}$Ca at 494 MeV. Three scattering
angles have been considered, namely 12.5$^{\circ}$, 18${^\circ}$ and
27${^\circ}$, corresponding to momentum transfers $q=1.2$, 1.7 and 2.5
fm$^{-1}$ at the quasielastic peak (QEP).

These experiments were prompted by the prediction \cite{Alb82}, based on the
$g'+\pi+\rho$ model of the residual particle-hole (ph) interaction, that the
attractive pion field should induce, at momentum transfers of the order of
$1\div2$ fm$^{-1}$, a softening and an enhancement of the isovector
spin-longitudinal ($\bbox{\sigma}\cdot\bbox{q}$) response ($R_L$), whereas a
quenching and a hardening should occur in the isovector spin-transverse
($\bbox{\sigma}\times\bbox{q}$) channel ($R_T$).

The original calculation, based on the random phase approximation (RPA) in a
simple infinite nuclear matter model, predicted a large $R_L/R_T$ ratio.
However, a first ($\vec{p},\vec{p}'$) experiment, performed at $q=1.7$
fm$^{-1}$ \cite{Car84,Ree86}, found a ratio consistent with, and even below,
one. Of course, ($\vec{p},\vec{p}'$) reactions do not allow the separation of
the isospin components in the response functions. As a consequence, the
interpretation of the above result was actually rather uncertain and model
dependent.

Charge-exchange reactions are not affected by this shortcoming and were
expected
to allow, in principle, a reliable mapping of the spin-isospin ph interaction
in nuclei. Hence, more ambitious ($\vec{p},\vec{n}$) experiments at three
different momentum transfers have been carried out \cite{McC92,Che93,Tad94},
still showing, however, no evidence for pion induced effects in the ratio
$R_L/R_T$, which is again around or below one.
On the other hand, the absence of any isoscalar contamination allows a
separation of the spin response functions on much firmer grounds, making a
direct comparison to the calculated responses more reliable.

A word of caution is in order here. The extraction of the responses from the
polarization observables is, strictly speaking, model dependent, since it
relies
on phenomenological $NN$ amplitudes and on a specific model for the treatment
of
distortion and absorption (for details on the extraction procedure see
Ref.~\cite{Che93}). While the first issue is not a severe one, because the
phenomenological $NN$ amplitudes give in fact a fairly good description of the
$^{2}$H data, the treatment of the reaction mechanism is more delicate, since
in Ref.~\cite{Che93} a simple effective number approximation has been employed
in performing the separation, defining the polarization observables as
combinations of response functions times the effective number of participating
nucleons,
\begin{equation}
  N_{\text{eff}} = \int d\bbox{b}\, T(b)e^{-\widetilde{\sigma}_{\text{tot}}}
    T(b),
\end{equation}
with
\begin{equation}
  T(b) = \int_{-\infty}^{+\infty} dz\,\rho(r=\sqrt{b^2+z^2}),
\end{equation}
$\rho(r)$ being the nuclear density and $\widetilde{\sigma}_{\text{tot}}$ the
effective $NN$ total cross section.
The latter is given by \cite{Smi89},
\begin{equation}
  \widetilde{\sigma}_{\text{tot}}(E) = \frac{2 m}{k} J_W/A,
\end{equation}
where the parametrization $J_W/A=0.6E$ MeV~fm$^3$ is adopted and $m$, $k$ and
$E$ are mass, momentum and kinetic energy of the projectile (actually, an
average over the incoming proton and outgoing neutron energies is taken);
$\widetilde{\sigma}_{\text{tot}}$ turns out to be $\approx26$ mb and the
effective number of neutrons $\approx2.3$ in $^{12}$C.
It should, however, be understood that this simply amounts to a
{\em definition} of the experimental response functions ``per nucleon'':
the latter, in principle, may contain contributions going beyond the effective
number treatment (those due to more sophisticated reaction mechanisms) and,
accordingly, the comparison to response functions extracted from
electromagnetic reactions might be questionable; yet, these hadronic responses
represent a legitimate experimental observable.

In Ref.~\cite{Tad94} the experimental response functions have been compared
to calculations in a model based on the distorted wave Born approximation
(DWBA)
with inclusion of RPA correlations \cite{Ich89}. The results there found are
surprising: the low values for $R_L/R_T$ appear to be due to a large excess of
strength in the transverse channel, up to a factor 2 larger than the DWBA
calculation, with or without RPA correlations, at the highest momentum.
In the spin-longitudinal channel one sees some crude agreement between the data
and the calculations, although the DWBA model performs too poorly to allow one
to draw definite conclusions about the presence of RPA effects. Indeed, the
model fails in reproducing both the momentum dependence of the strength and
the position of the QEP.

It is the aim of the present letter to show how a model, developed in
Ref.~\cite{DeP93}, based on Glauber theory and accounting for certain classes
of many-body correlations, is able to give a fairly good description of the
spin-longitudinal data, hinting at some evidence of pionic RPA effects.
Moreover, while the spin-transverse data remain still unexplained, we show how
the two-step mechanism, in spite of its sizable contribution in this channel,
is insufficient to provide an explanation of the discrepancy.

We now briefly sketch the formalism; for further details the reader is referred
to Ref.~\cite{DeP93}. The basic quantity in the calculation of the nuclear
response functions is the polarization propagator, which reads
\begin{eqnarray}
  \Pi_\alpha(\bbox{q},\bbox{q}';\omega) &=&
    \sum_{n\ne0}\langle\psi_0|{\hat O}_\alpha(\bbox{q})|\psi_n\rangle
    \langle\psi_n|{\hat O}^\dagger_\alpha(\bbox{q}')|\psi_0\rangle
  \nonumber\\
  &&\quad\times\left[\frac{1}{\hbar\omega-(E_n-E_0)+i\eta}
    -\frac{1}{\hbar\omega+(E_n-E_0)-i\eta}\right],
\end{eqnarray}
where \{$|\psi_n\rangle$\} is a complete set of nuclear eigenstates of energy
$E_n$ and ${\hat O}_\alpha(\bbox{q})$ the second quantized expression of the
vertex operator. In the case of the spin-isospin modes one has
\begin{mathletters}
  \label{eq:OLT}
\begin{eqnarray}
  O_L(\bbox{q},\bbox{r})&=&\tau_\alpha
  \bbox{\sigma}\cdot\hat{\bbox{q}}
  e^{i\bbox{q}\cdot\bbox{r}} \\
  O_T(\bbox{q},\bbox{r})&=&\frac{\tau_\alpha}{\sqrt{2}}
  \bbox{\sigma}\times\hat{\bbox{q}}
  e^{i\bbox{q}\cdot\bbox{r}}
\end{eqnarray}
\end{mathletters}
in the {\em longitudinal} and {\em transverse} channels, respectively.

After an angular momentum expansion, the responses are defined as
\begin{equation}
  R_{L,T}(q,\omega) =
    -\frac{1}{4\pi^2}\text{Im}\sum_J(2J+1)\Pi_{J(L,T)}(q,q;\omega),
\end{equation}
where $\Pi_{J(L,T)}$ are obtained in the continuum RPA by solving a set of
coupled integral equations, using Woods-Saxon wave functions for the ph states
and the $g'+\pi+\rho$ model for the ph interaction (in the calculations we
present below, $g'=0.7$ has been used).

Another important feature of the nuclear dynamics is represented by the {\em
spreading width} of the ph states, i.~e. by their coupling to more complicated
configurations. We incorporate this element in a phenomenological manner, by
including in the mean field polarization propagator a ph self-energy, designed
to fit the empirical particle widths \cite{Mah81} (see also Ref.~\cite{Smi88}).
Note that it turns out to be rather easy to accomodate the spreading width in
our scheme, since all calculations are performed in momentum space, although we
are dealing with finite systems.

Another relevant aspect of our approach is the reaction mechanism.
This has been implemented in the framework of the Glauber theory \cite{Gla59},
including one- and two-step terms: the one-step contribution is obtained by
substituting the vertex operators (\ref{eq:OLT}) with
\begin{equation}
  O^{\text{surf}}_{L,T}(\bbox{q},\bbox{r}) = \frac{1}{(2\pi)^2 f_{L,T}(q)}
    \int d\bbox{b}\,d \bbox{\lambda}\,
    e^{-\widetilde{\sigma}_{\text{tot}} T(b)/2}
    e^{i(\bbox{q}-\bbox{\lambda})\cdot\bbox{b}}f_{L,T}(\lambda)
    O_{L,T}(\bbox{\lambda},\bbox{r}),
\end{equation}
where $f_{L,T}$ are the elementary isovector spin-longitudinal and
spin-transverse $NN$ scattering amplitudes.

The two-step contribution has been evaluated through the convolution of two
free response functions \cite{Smi87}, i.~e.
\begin{eqnarray}
  R^{(2)}_{L,T}(q,\omega) &=&
  \frac{{\cal D}_2}{k^2}\frac{2}{|f_{L,T}(q)|^2}
  \nonumber\\
  &&\quad\times\int d\bbox{q}' \int_0^\omega d\omega'
  \big|f_{L,T}(q')\big|^2 R_{L,T}(q',\omega')
  \big|f_{00}(|\bbox{q}-\bbox{q}'|)\big|^2
     R_{00}(|\bbox{q}-\bbox{q}'|,\omega-\omega'),
\end{eqnarray}
where $R_{00}$ and $f_{00}$ are the scalar-isoscalar nuclear response and $NN$
amplitude, respectively, and
\begin{equation}
  {\cal D}_2=\frac{1}{2}\int d\bbox{b}\,T^2(b)
  e^{-\widetilde{\sigma}_{\text{tot}}T(b)}
\end{equation}
is connected to the effective number of pairs participating in the double
scattering process.

\begin{figure}[p]
\begin{center}
\mbox{\epsfig{file=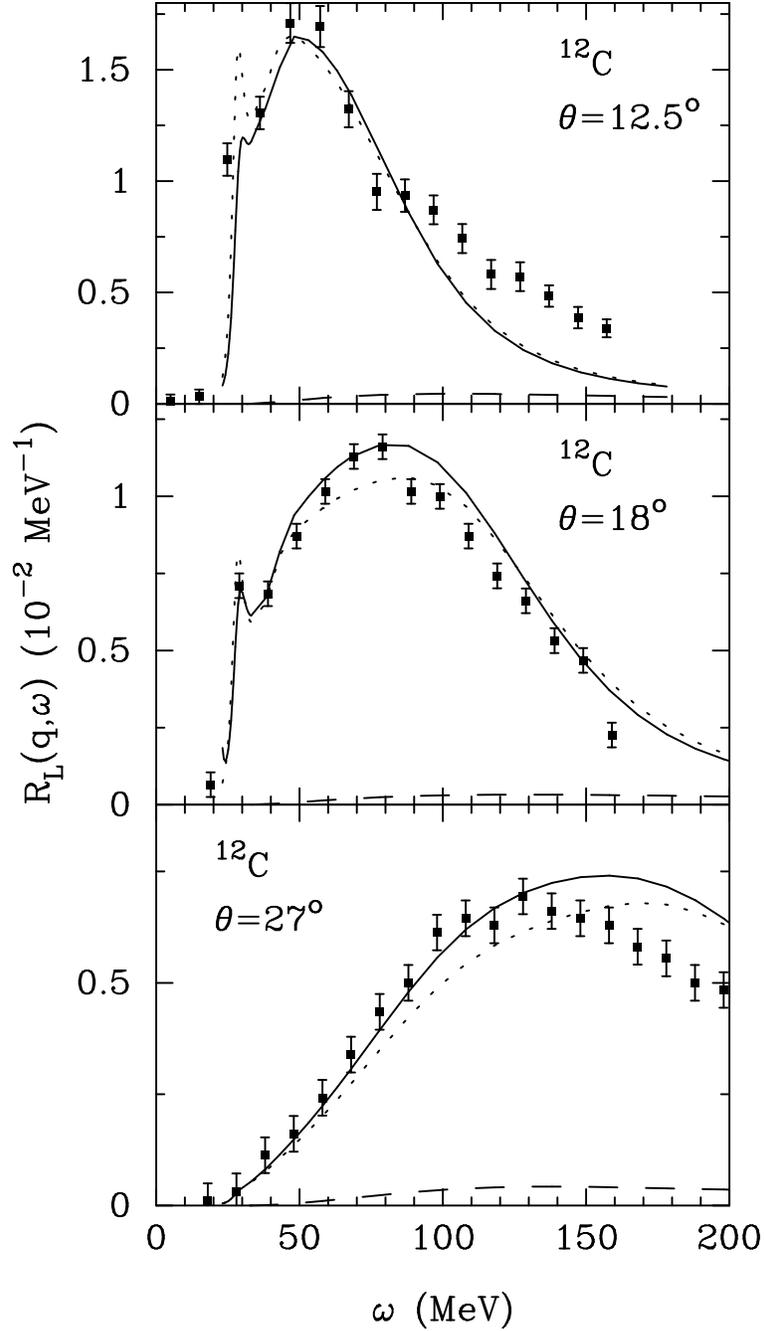}}
\caption{ Spin-longitudinal isovector responses for ($\vec{p}$,$\vec{n}$)
reactions at 494 MeV, with (solid) and without (dotted) RPA correlations.
The two-step contribution is also displayed separately (dashed). Data are from
Ref.~\protect{\cite{Tad94}}.
  }
\label{fig:fig1}
\end{center}
\end{figure}

In Fig.~\ref{fig:fig1} the calculated spin-longitudinal responses are compared
to the data. It is immediately apparent that the model gives a good description
of the data up to 100$\div$150 MeV of excitation energy, depending on the
angle.

At momentum transfers around 1.2 fm$^{-1}$ ($\theta=12.5^{\circ}$) the
longitudinal ph interaction is rather small and both the uncorrelated and RPA
results well reproduce the QEP shape and strength.
At $\theta=18^{\circ}$ the RPA calculation works quite well on the left side of
the QEP, whereas on the right the RPA and free responses are very close and
slightly overestimate the strength.
At $\theta=27^{\circ}$, again the RPA model is quite accurate on the left of
the
QEP, whereas on the right the data show a faster decrease than predicted by
both
the RPA and the uncorrelated calculations.
The two-step contribution is very small at the lowest angle and $\sim$5\% of
the
total at $\theta=27^{\circ}$.
\begin{figure}[p]
\begin{center}
\mbox{\epsfig{file=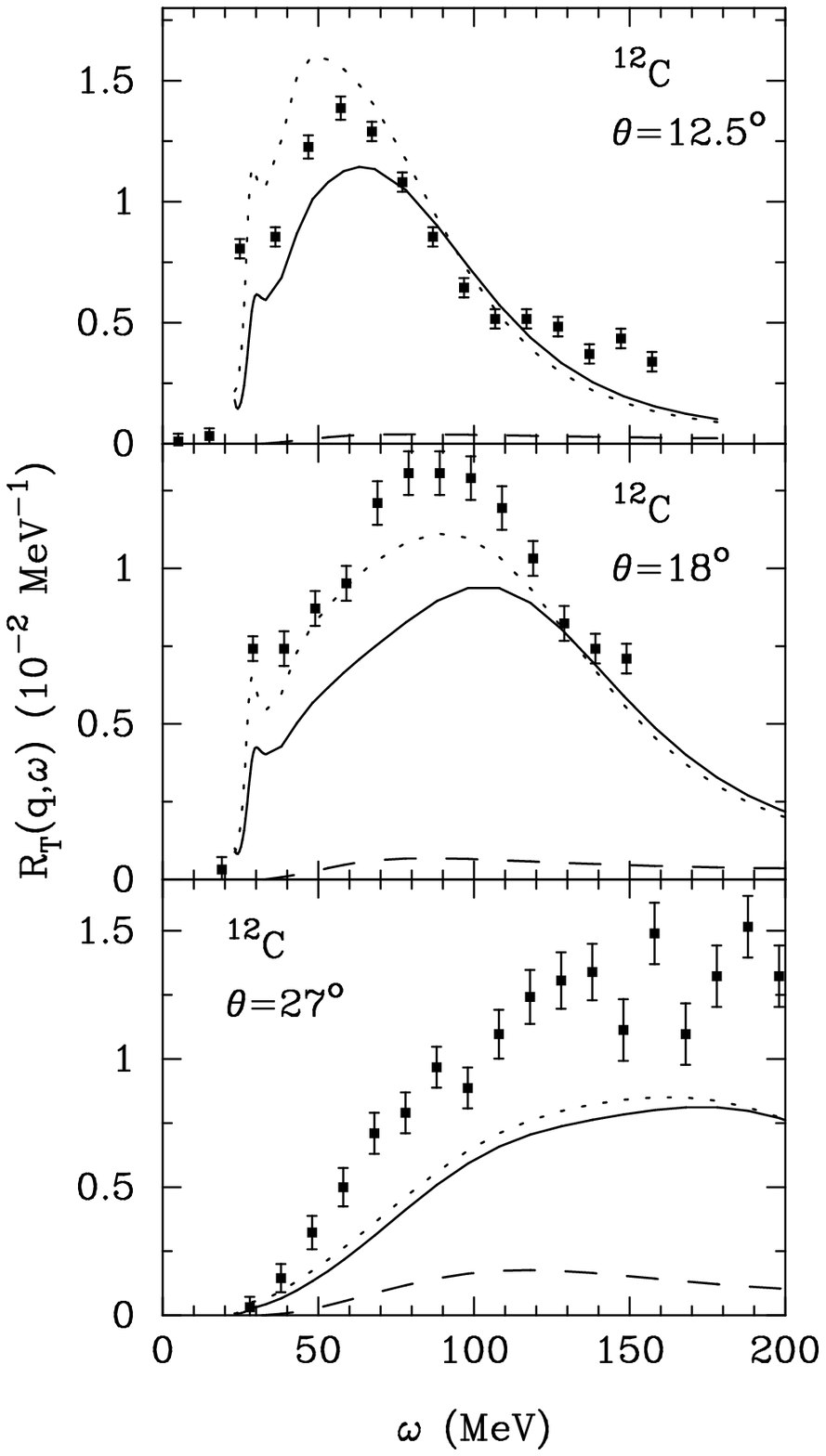}}
\caption{ Spin-transverse isovector responses for ($\vec{p}$,$\vec{n}$)
reactions at 494 MeV. The meaning of the lines is explained in
Fig.~\protect{\ref{fig:fig1}}. Data are from Ref.~\protect{\cite{Tad94}}.
  }
\label{fig:fig2}
\end{center}
\end{figure}

It should be noted that the low energy part of the spectra is where the present
model is expected to be more reliable, since other aspects of the nuclear
dynamics not included here (two-particle--two-hole (2p2h), meson production,
...) are of importance at higher energies. Moreover, at the highest momenta
effects from relativistic kinematics start to be appreciable
(at $q=2.5$ fm$^{-1}$ the QEP position is shifted downward of $\approx8$ MeV).

Note also that an important role in yielding the right shape to the
quasielastic
response is played by the spreading width of the ph states: as discussed in
Ref.~\cite{DeP93}, the effect of the latter amounts to redistribute the
strength around the QEP, increasing the response in the high and low energy
regions, while reducing it at the peak. A much smaller, but still appreciable,
increase of the response in the high and low energy tails is also due to the
fact that the calculations have been performed at fixed scattering angle, which
accounts for the dependence of the momentum transfer on the energy loss.

Next, we consider the spin-transverse responses, displayed in
Fig.~\ref{fig:fig2}. Here the situation is rather surprising: the data show,
in fact, a large excess of strength in this channel, which is the same entering
in magnetic ($e$,$e'$) scattering. In the latter case, the data point to a
quenching of the free response, which is indeed obtained in models based on
the $g'+\pi+\rho$ residual interaction \cite{Alb86}.

The same quenching is also present in the RPA responses displayed in
Fig.~\ref{fig:fig2}, in strong contradiction with the data.
In this channel the two-step contribution is much more sizable, about
20$\div$30\% of the ph response at $\theta=27^{\circ}$, but still unable to
fill
the gap. Describing the data within a RPA framework would require a ph
interaction radically different from the one currently accepted; on the other
hand, it is difficult to imagine that dynamical effects such as 2p2h
excitations should play a role comparable to, and even larger than, the one
they have in ($e$,$e'$) scattering, since in the present case the incident
proton actually probes a lower nuclear density.

{}From the analysis of Figs.~\ref{fig:fig1} and \ref{fig:fig2}, we derive the
following observations:

The spin-longitudinal response is well described in the energy domain where a
ph
based model can be expected to be reliable. Note that all the ingredients
entering into the calculations are of importance and none can be neglected.
This is true over a wide range of momentum transfers; in particular, when the
pion driven ph force is felt, then the RPA correlations remarkably improve the
description. Of course, before drawing any definite conclusion the effect of
all the elements not included in the present model, such as $\Delta$ degrees
of freedom, 2p2h, ..., should be assessed.

On the other hand, the excess of strength in the spin-transverse response is
unexplained. As we have seen, multiple scattering is sizable and cannot be
neglected, but it is not enough to fill the gap. However, owing to the accuracy
achieved in the description of the longitudinal response, it is very unlikely
that the discrepancy found in the transverse channel arises from uncertainties
in the reaction mechanism (in addition, one should note that the model gives a
good description of the ($p$,$p'$) cross sections as well \cite{DeP93}).
This excess of transverse strength is probably also responsible for the
excess of cross section found in the ($p$,$n$) reaction at 800 MeV
\cite{DeP93}.

On the theoretical side, it would be interesting to understand the poor
performance of the DWBA calculations \cite{Tad94}, since the results obtained
in this model are expected to coincide, in the eikonal limit, with those
obtained in the Glauber approximation.

\end{document}